\documentclass[twocolumn,twoside]{revtex4}
\pdfoutput=1
\usepackage{graphicx}
\usepackage{fancyhdr}
\usepackage{hepnames}
\newcommand{\GeV}{\ensuremath{\,\text{Ge\hspace{-.08em}V}}\xspace}

\newcommand{\abinv} {\mbox{\ensuremath{\,\text{ab}^\text{$-$1}}}\xspace}
\newcommand{\fbinv} {\mbox{\ensuremath{\,\text{fb}^\text{$-$1}}}\xspace}

\begin{document}

\title{First look at time-dependent CP violation using early Belle II data}

%

\author{Stefano Lacaprara\\
on behalf of the Belle II Collaboration}
\affiliation{INFN sezione di Padova}

\begin{abstract}
    Time dependent CP-violation phenomena are a powerful tool to
    precisely measure fundamental parameters of the Standard Model and search
    for New Physics.
    The Belle II experiment at the SuperKEKB energy-asymmetric $\Pep\Pem$
    collider is a substantial upgrade of the B factory facility at the Japanese
    KEK laboratory. The design luminosity of the machine is
    $8\times{10}^{35}~cm^{-2}s^{-1}$ and the Belle II experiment aims to record
    $50~ab^{-1}$ of data, a factor of 50 more than its predecessor. From
    February to July 2018, the machine has completed a commissioning run,
    achieved a peak luminosity of $5.5\times10^{33}~cm^{-2}s^{-1}$ , and Belle
    II has recorded a data sample of about $0.5~fb^{-1}$ . 
    
    Main operation of SuperKEKB has started in March 2019. This early data set
    is used to establish the performance of the detector in terms of
    reconstruction efficiency of final states of interest for the measurement
    of time dependent CP violation, such as $\PJpsi\PKz$ , $\Petaprime \PKz$,
    and $\phi \PKz$.  A first assessment of the B flavor tagging capabilities of
    the experiment will be given, along with estimates of the Belle II
    sensitivity to the CKM angles $\phi_1/\beta$ and $\phi_2/\alpha$ and to
    potential New Physics contributions in penguin amplitudes dominated decays
    and in $b\to s\gamma$ transitions.

    In this talk we will present estimates of the sensitivity to $\phi_1$ in the
    golden channels $\Pqb\to\Pqc\APqc\Pqs$ 
    and in the penguin-dominated modes 
    $\PBz\to\Petaprime\PKz,\quad\Pphi\PKz,\quad \PKz\Pgpz(\gamma)$.
    A study for the time-dependent analysis of $\PBz\to\Pgpz\Pgpz$,
    relevant for the measurement of $\phi_2$, and feasible only in the clean
    environment of an $\Pep\Pem$ collider, will also be given.

\end{abstract}

\maketitle

\thispagestyle{fancy}


\section{Introduction}
The Belle II experiment~\cite{Abe:2010gxa} is currently taking data at
SuperKEKB~\cite{10.1093/ptep/pts083} $\Pep\Pem$ collider at KEK, in Tsukuba, Japan.
Its main goal is to search for physics beyond the Standard Model (SM) in \PB,
\PD, and \Ptau decays, performing precise measurements of quantities for which
the SM provide reliable calculation or looking for exotic particles and decays.
The SuperKEKB asymmetric collider works at a center of mass energy corresponding (or close
    to) the \PUpsilonFourS resonance. The target integrated luminosity is
$50~ab^{-1}$ in about five year of operation, with a ultimate instantaneous
luminosity of $8\times10^{35}~cm^{-2}s^{-1}$, 40 times larger than that of
KEKB, its predecessor.

Belle II has collected an integrated luminosity of about 500~pb$^{-1}$ in year
2018 (Phase 2) with an incomplete detector, where only a small fraction of the
silicon vertex sub-detector was installed.
Starting from march 2019, Belle II is collecting more data at \PUpsilonFourS
center of mass energy, with all sub-detectors fully installed and operational.

One of the major goal of the Belle II physics program~\cite{Kou:2018nap} is to
improve the measurement of the CP violating parameter of
Cabibbo-Kobayashi-Maskawa (CKM) matrix, via the time-dependent CP violation
(TDCPV) technique in the \PB sector.

In this contribution, we will present the expected sensitivity on CKM Unitary
Triangle (UT) angles $\phi_1/\beta$, both in tree- and penguin-mediated
transition, and $\phi_2/\alpha$.
Additionally, searches for New Physics using TDCPV techniques in the
$\PKz\Pgpz(\gamma)$ decay will be described.

\section{Belle II detector}
The Belle II detector is major upgrade of the previous Belle one for almost all subsystem.
Particularly relevant for the TDCPV analysis is the new vertex detector, and
the particle identifier (PID) ones.

The vertex detector comprises two layers of silicon pixel (PXD), the closest
placed at $r=14~mm$ from the beam line, and four layers of silicon strips (SVD).
In spite of the reduced boost of the \PB particle at SuperKEKB with respect to
KEKB ($\beta\gamma=0.28$ vs $0.45$), an improvement of secondary vertex
precision, and so of $\Delta{t}$ of about 30\% is expected.
The tracking system is completed by a central drift chamber (CDC), installed
inside a solenoid providing a 1.5~T magnetic field.
The CDC is new, larger than the one used at Belle, 
with smaller cell size and longer lever arm. 

Two new PID detectors have been installed to provide separation between charged
hadrons (mostly \Pgp and \PK). In the barrel region, the Imaging Time Of Propagation
(iTOP) detector is used, and the Aerogel Cherenkov (ARICH) covers the forward
region. Both are based on the Cherenkov effect. The iTOP uses precise determination
of time and position of arrival of Cherenkov photons to an array of Micro
Channel Plate Photo Multiplier Tube. The photons are produced by charged
particle crossing 2 cm of finely polished quartz.
The ARICH uses an aerogel radiator with focusing capability, and collect
Cherenkov photons via Hybrid Avalanche Photo Detector.  The CDC provide
additional information on PID measuring the specific ionization energy deposit.
The overall performances are expected to provide an efficiency for
identification of \Pgp (\PK) greater than $90\%$ with a mis-identification rate
lower than $10\%$ up to $4~\GeV/c$ for particle momentum.

The improved PID, together with better algorithm based on multivariate
selectors, is expected to increase the flavour tagging effective efficiency to
$\sim37\%$, to be compared to $\sim30\%$ obtained at Belle.

\section{Measurement of $\phi_1/\beta$}
The most precise determination of $\sin2\beta$ is obtained from TDCPV analysis
of tree mediated $\Pqb\to\Pqc$ processes, dominated by the decay $\PBz\to\PJpsi\PKz$.
An improved measurement might indicate New Physics in case of incompatibility
with other measurement of the UT parameters.
The measurement of $\sin2\beta$ in $\PBz\to\Pqc\APqc\PKz$ decays will be soon
dominated by systematics uncertainties, thanks to the large dataset expected
at Belle II. Dominant irreducible systematics come from alignment of vertex
detector, $\Delta{t}$ resolution, and tag-side interference.

\begin{figure}[htb]
\centering
\includegraphics[width=80mm]{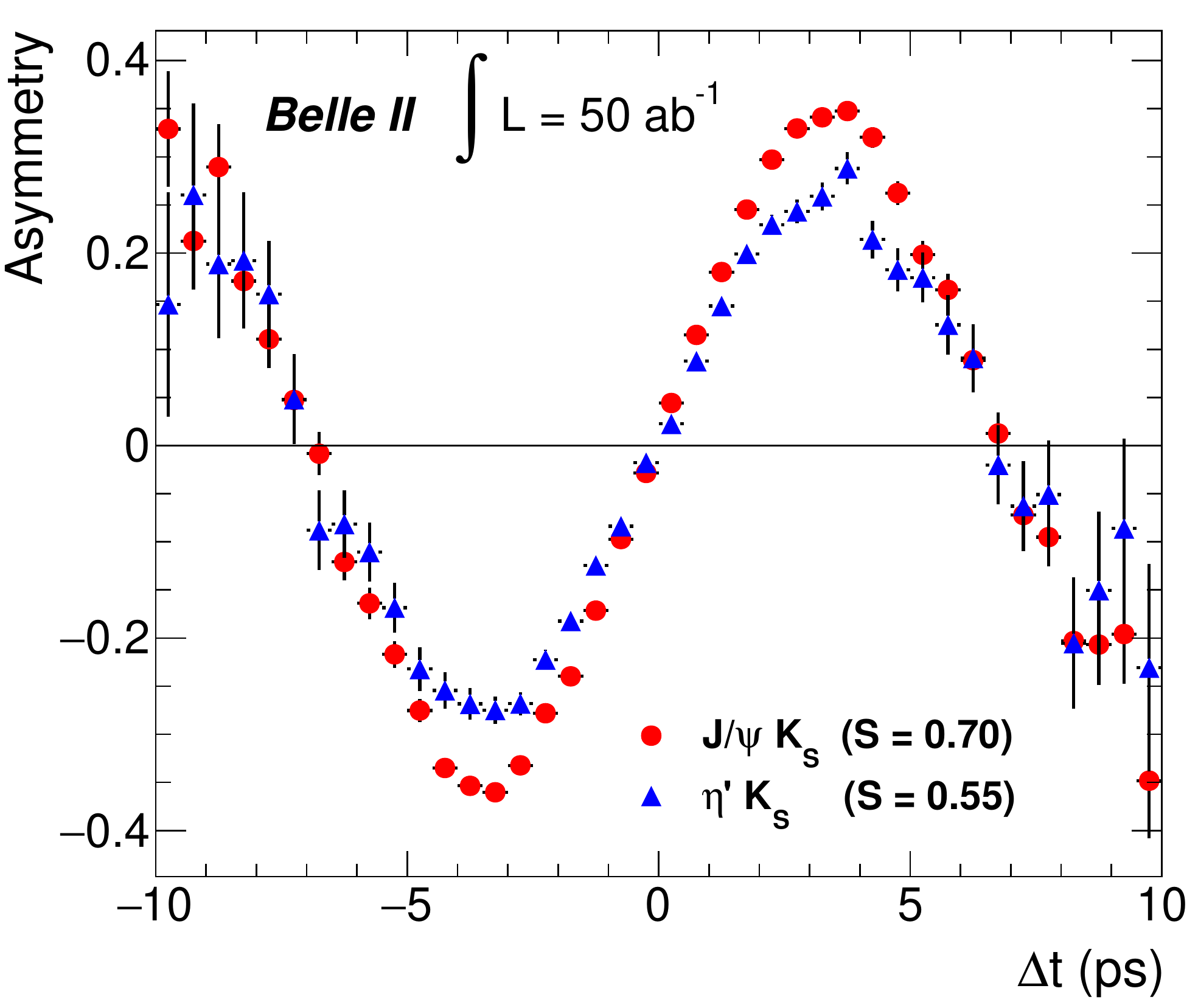}
\caption{Simulated time dependent CP asymmetry in the $\PJpsi\PKz$ (red dots) and
    $\Petaprime\PKz$ (blue triangles) expected with an integrated luminosity of
    50\abinv, using $S=0.70(0.55)$ for the former (latter) channel, respectively.} \label{Fig:Asym}
\end{figure}

Independent determination of time dependent $CP$ asymmetry $S$ will come from
penguin dominated  $\Pqb\to\Pqs\Pquark\APquark$  ($\Pquark=\Pqs,\Pqd,\Pqu$).
These decays are suppressed with respect to  $\Pqb\to\Pqc$, due to the
presence of loop amplitudes, but New Physics can introduce new phases in the
loop, and shift the value of $S$ as measured in the tree-dominated diagram.
A significant difference would be a clear indication of New Physics, as shown
in Fig.~\ref{Fig:Asym}.

The expected sentitivity for several final states have been studied: the most
interesting are $\PBz\to\Petaprime\PKz$ and $\Pphi\PKz$.
The former has many different final states, from the $\Petaprime$ decay, all
including neutrals. The latter have final states with \PK and \Pgp, and the PID
performances will be critical.
These two channels are also the cleaner from the thoerotical point of view, so
the comparison of $S$ to that of tree-dominated diagram is more sensitive.

Given the suppressed amplitudes, the precision of $S$ in most of the decay
channels is expected to be dominated by statistical uncertainties. A
notable exception is the decay $\PBz\to\Petaprime\PKz$, where the systematics
uncertainties are expected to be similar to the statistical one around
$10-20\abinv$.

To probe for New Physics, it is also interesting to look for TDCPV when none
is expected in the SM. An example is the decay channel
$\PBz\to\PKs\Ppizero\Pgamma$, which includes many neutrals in the final states,
making very hard, if possible at all, for LHCb. The decay
$\Pqb\to{\Pqs\gamma_R}$ is helicity suppressed with respect to the analogous
with $\gamma_L$, but New Physics can enhance the $b\to{s\gamma_R}$ decay rate.
The interference between $\PBz\to{f_{CP}}\gamma_R$ and
$\PBz\to\PaBz\to{f_{CP}}\gamma_R$ happens only for helicity suppressed decay,
and a measurement of time dependent asymmetry $S$ in this channel larger than
SM prediction (few \%~\cite{PhysRevD.75.054004}) would be an evidence of New Physics.

A summary of expected precision on time dependent ($S$)
and direct ($A$) CP violation parameters is presented in Table~\ref{tab:phi1}.

\begin{table}[ht]
\begin{center}
\caption{Expected uncertainties on time dependent ($S$)
    and direct ($A$) CP violation parameters for integrated luminosity of 5 and 50\abinv. Current World Average (WA) uncertainties are also reported~\protect\cite{Amhis:2016xyh}}.
\begin{tabular}{lrrrrrr}
    & \multicolumn{2}{c}{\textbf{WA}} &  \multicolumn{2}{c}{\textbf{5\abinv}} &  \multicolumn{2}{c}{\textbf{50\abinv}} \\
    \textbf{Channel} & \textbf{$\sigma(S)$} & \textbf{$\sigma(A)$} & \textbf{$\sigma(S)$} & \textbf{$\sigma(A)$} & \textbf{$\sigma(S)$} & \textbf{$\sigma(A)$} \\
    \hline
            $J/\psi K^0$             & 0.22 & 0.21 & 0.012 & 0.011 & 0.0052 & 0.0090 \\
            $\eta^{\prime} K^0$      & 0.06 & 0.04 & 0.032 & 0.020 & 0.015 & 0.008 \\
            $\phi K^0 $              & 0.12 & 0.14 & 0.048 & 0.035 & 0.020 & 0.011 \\
            $\omega K^0_S$           & 0.21 & 0.14 & 0.08  & 0.06  & 0.024 & 0.020 \\
            $K^0_S \pi^0 \gamma$     & 0.20 & 0.12 & 0.10  & 0.07  & 0.031 & 0.021 \\
            $K^0_S \pi^0$            & 0.17 & 0.010 & 0.09 & 0.06  & 0.028 & 0.018 \\
    
\end{tabular}
\label{tab:phi1}
\end{center}
\end{table}

\section{Measurement of $\phi_2/\alpha$}

The second UT angle $\phi_2/\alpha$ can be measured using the decay
$\PBz\to\Pgp\Pgp$ as well as $\Prho\Prho$, using an isospin analysis~\cite{PhysRevLett.65.3381},
due to the fact that there are two amplitudes of comparable size but different weak phase.
The isospin techniques requires to determine the magnitude and phases of
$\PBz\to\Pgpp\Pgpm$ as well as  $\PBz\to\Pgpz\Pgpz$. Previous B-factories
measurements include both for $\PBz\to\Pgp\Pgp$, but the phase of
$\PBz\to\Pgpz\Pgpz$ has not been measured, resulting in a eightfold ambiguity in
the determination of $\phi_2$.
By measuring the phase, namely the time dependent asymmetry of
$\PBz\to\Pgpz\Pgpz$, the ambiguity can be reduced by a factor of two or four.
Given the neutral final state, this decay channel is very difficult to study
at LHCb.

\begin{figure}[tb]
\centering
\includegraphics[width=80mm]{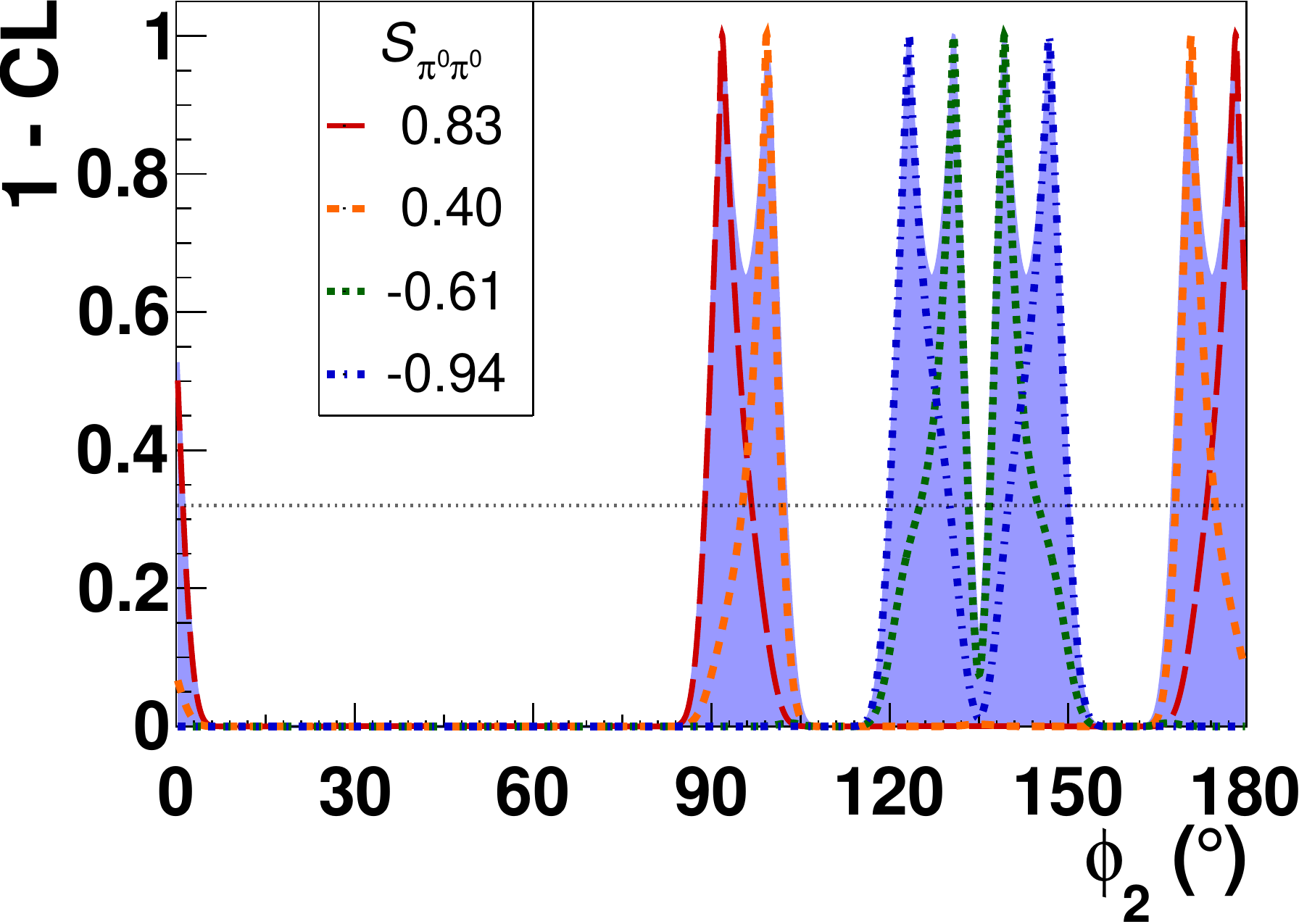}
\caption{Expected CL-1 for $\phi_2$, using isospin analysis for
    $\PBz\to\Pgp\Pgp$ final state, only with improvement on current measurement
    (blue area) and including also the determination of $S_{\Pgpz\Pgpz}$ values
    (colored lines) for four different central value of $S$.
    } \label{Fig:phi2}
\end{figure}

At Belle II, the high integrated luminosity allows to study TDCPV for
$\PBz\to\Pgpz\Pgpz$ using the Dalitz decays $\Pgpz\to\Pphoton\Pep\Pem$ and the
$\Pphoton\to\Pep\Pem$ conversion on beam pipe and first layers of vertex
detector.
The \PBz decay vertex can be reconstructed from the \PBz flight direction and
that of $\Pep\Pem$, with a vertex resolution only 50\% worse than that obtained
with charged final state, such as  $J/\psi K^0$. With the full integrated
luminosity of 50\abinv, Belle II expects to reconstruct $\sim270$
$\PBz\to\Pgpz\Pgpz$ signal events with Dalitz decay, and $\sim50$ with a $\Pphoton$ conversion.
The foreseen precision on $S$ would be $\sim0.28$, and the impact on reducing
the ambiguity on $\phi_2$ is shown in Fig.~\ref{Fig:phi2} for four different
possible values of $S_{\Pgpz\Pgpz}$.

The $\PBz\to\Pgp\Pgp$ will provide a resolution on $\phi_2$ around $2^\circ$:
combining with the similar isospin analysis of $\Prho\Prho$ final state, the
ultimate precision will be $0.6^\circ$, using the full statistics.

\section{Conclusion and outlook}

The Belle II experiment is taking data at SuperKEKB in Tsukuba, Japan. In the
physics program of the experiment, the study of time dependent CP violation will play a major role.
The ultimate precision on $\phi_1$ and $\phi_2$ is expected to go below
$\sim1^\circ$, with interesting prospect in term of New Physics searches.

Looking at B-factory (Belle and BaBar) past history, with an integrated
luminosity around 10\fbinv, Belle II could re-discover TDCPV in the
$\PBz\to\PJpsi\PKs$ final state, as well the decay into $\Petaprime\PKs$.
Larger dataset, around 40\fbinv would be required for TDCPV in the latter
channel as well as in the $\PBz\to\Pgpp\Pgpm$ one.
Even larger $\sim100\fbinv$ were needed for TDCPV in ${\Pphi\PKs}$.
For channel  $\PBz\to\PKs\Ppizero\Pgamma$, Belle II can provide interesting
improvement with few \abinv, while the full dataset of 50\abinv will be needed
for the challenging $\PBz\to\Pgpz\Pgpz$.

More information about the Belle II physics program can be found on~\cite{Kou:2018nap}.

\bigskip 
\bibliography{lacaprara}

\begin{thebibliography}{6}
\expandafter\ifx\csname natexlab\endcsname\relax\def\natexlab#1{#1}\fi
\expandafter\ifx\csname bibnamefont\endcsname\relax
  \def\bibnamefont#1{#1}\fi
\expandafter\ifx\csname bibfnamefont\endcsname\relax
  \def\bibfnamefont#1{#1}\fi
\expandafter\ifx\csname citenamefont\endcsname\relax
  \def\citenamefont#1{#1}\fi
\expandafter\ifx\csname url\endcsname\relax
  \def\url#1{\texttt{#1}}\fi
\expandafter\ifx\csname urlprefix\endcsname\relax\def\urlprefix{URL }\fi
\providecommand{\bibinfo}[2]{#2}
\providecommand{\eprint}[2][]{\url{#2}}

\bibitem[{\citenamefont{Abe et~al.}(2010)}]{Abe:2010gxa}
\bibinfo{author}{\bibfnamefont{T.}~\bibnamefont{Abe}} \bibnamefont{et~al.}
  (\bibinfo{collaboration}{Belle-II}) (\bibinfo{year}{2010}),
  \eprint{1011.0352}.

\bibitem[{\citenamefont{Ohnishi et~al.}(2013)}]{10.1093/ptep/pts083}
\bibinfo{author}{\bibfnamefont{Y.}~\bibnamefont{Ohnishi}} \bibnamefont{et~al.},
  \bibinfo{journal}{PTEP} \textbf{\bibinfo{volume}{2013}}
  (\bibinfo{year}{2013}), ISSN \bibinfo{issn}{2050-3911}.

\bibitem[{\citenamefont{Kou et~al.}(2018)}]{Kou:2018nap}
\bibinfo{author}{\bibfnamefont{E.}~\bibnamefont{Kou}} \bibnamefont{et~al.}
  (\bibinfo{collaboration}{Belle-II}) (\bibinfo{year}{2018}),
  \eprint{1808.10567}.

\bibitem[{\citenamefont{Ball et~al.}(2007)\citenamefont{Ball, Jones, and
  Zwicky}}]{PhysRevD.75.054004}
\bibinfo{author}{\bibfnamefont{P.}~\bibnamefont{Ball}},
  \bibinfo{author}{\bibfnamefont{G.~W.} \bibnamefont{Jones}}, \bibnamefont{and}
  \bibinfo{author}{\bibfnamefont{R.}~\bibnamefont{Zwicky}},
  \bibinfo{journal}{Phys. Rev. D} \textbf{\bibinfo{volume}{75}},
  \bibinfo{pages}{054004} (\bibinfo{year}{2007}),
  \urlprefix\url{https://link.aps.org/doi/10.1103/PhysRevD.75.054004}.

\bibitem[{\citenamefont{Amhis et~al.}(2017)}]{Amhis:2016xyh}
\bibinfo{author}{\bibfnamefont{Y.}~\bibnamefont{Amhis}} \bibnamefont{et~al.}
  (\bibinfo{collaboration}{HFLAV}), \bibinfo{journal}{Eur. Phys. J.}
  \textbf{\bibinfo{volume}{C77}}, \bibinfo{pages}{895} (\bibinfo{year}{2017}),
  \eprint{1612.07233}.

\bibitem[{\citenamefont{Gronau and London}(1990)}]{PhysRevLett.65.3381}
\bibinfo{author}{\bibfnamefont{M.}~\bibnamefont{Gronau}} \bibnamefont{and}
  \bibinfo{author}{\bibfnamefont{D.}~\bibnamefont{London}},
  \bibinfo{journal}{Phys. Rev. Lett.} \textbf{\bibinfo{volume}{65}},
  \bibinfo{pages}{3381} (\bibinfo{year}{1990}),
  \urlprefix\url{https://link.aps.org/doi/10.1103/PhysRevLett.65.3381}.

\end{thebibliography}


\begin{thebibliography}{9}   
\end{thebibliography}

\end{document}